\documentclass[]{llncs}
\usepackage{amsfonts}
\usepackage{amsmath}
\usepackage{amssymb}
\usepackage{graphics}
\usepackage[dvips]{epsfig}
\usepackage{times}
\usepackage{algorithm}
\usepackage{algorithmic}
\usepackage{fancyhdr}
\usepackage{ifthen}
\usepackage[english]{babel}
\newcommand\settitle[2][]{%
 \title{#2}
 \ifthenelse{\equal{#1}{}}%
  {\fancyhead[RO]{\nouppercase #2 \qquad \thepage}}%
  {\fancyhead[RO]{\nouppercase #1 \qquad \thepage}}%
}

\fancypagestyle{plain}{%
\fancyhf{}

}

   \newtheorem{ex}{Example}
   \newtheorem{Bemerkung}{Remark}

   \newcommand{\D}{\displaystyle}

   \def\C{{\mathbb C}}
   \def\R{{\mathbb R}}

   \def\Pf{{\it Proof.$\;\;$}}

   \def\qed{\hfill$\diamond$}

   \def\cA{{\mathcal A}}
   \def\cM{{\mathcal M}}
   
   \def\cS{{\mathcal S}}
   \def\cV{{\mathcal V}}

\def\C{{\mathbb C}}

\def\M{{\mathbb M}}

\def\R{{\mathbb R}}

\def\Pf{{\it Proof.$\;$}}

\def\qed{\hfill$\blacksquare$}

\def\diag{\mbox{\rm diag}}

\def\Pr{\mbox{\rm Pr}}

\def\span{\mbox{\rm span}}

\def\({\langle}
\def\){\rangle}
\def\mb{\boldsymbol}
\def\im{{\rm i}}

\def\cA{{\mathcal A}}

\def\cG{{\mathcal G}}
\def\cH{{\mathcal H}}

\def\cM{{\mathcal M}}

\def\cS{{\mathcal S}}

\def\cV{{\mathcal V}}

\def\1{\mb1}

\def\v0{{\bf 0}}

\def\ve{{\bf e}}

\def\vs{{\bf s}}

\def\vx{{\bf x}}
\def\vy{{\bf y}}
\def\vz{{\bf z}}

\def\vE{{\bf E}}

\def\vG{{\bf G}}

\def\ov{\overline}

\begin{document}

\settitle[Bell Inequality]{Bell's Inequality and Heisenberg Measurements \\on Relativistic Quantum Systems\\ 
\medskip
({\rm Preprint, 25 April, 2023)}}

\author{Ulrich Faigle}
\institute{Mathematisches Institut\\
           Universt\"at zu K\"oln\\
           Weyertal 80, 50931 K\"oln, Germany\\
\email{faigle@zpr.uni-koeln.de}\\[1ex]}

\maketitle

\thispagestyle{plain}
\begin{abstract}
Bell's inequality plays an important role with respect to the Einsteinian question about the physical reality of quantum theory. While Bell's inequality is usually viewed within the geometric framework of a Hilbert space quantum model, the present note extends the theory of Heisenberg measurements to quantum systems with representations in general orthogonal geometric spaces and, in particular, the Minkowski spaces of relativity theory. A Feynmanian numerical example exhibits two measurements that admit a joint probabilistic interpretation in Min\-kowski space while they are not jointly observable in Hilbert space.

\medskip
The analysis shows that probabilistic interpretations of quantum measurements may depend not only on the measuring instruments and the system states but also on the geometric space in which the measurements are conducted.  In particular, an explicit numerical example is given of a Heisenberg measurement with a complete set of common observables that violates Bell's inequality in Minkowski space but, {\it mutatatis mutandis}, satisfies it in Hilbert space.
\end{abstract}

\section{Introduction}\label{sec:Introcduction}
In the well-known EPR {\it Gedankenexperiment}, Einstein, Rosen and Podolsky~\cite{EPR} question the world-view of quantum theory because of missing \emph{hidden states} that would imply deterministic properties of particles to be measured. Assuming that propagation of information cannot exceed the speed of light, a particle should be in a well-defined state, whether it is measured or not. According to quantum theory, this \emph{\it local realism} may not always be guaranteed. In order to decide the issue, Bell~\cite{Bell64,Bell66} devised an inequality and an  experiment which quantum theory predicts to violate, Einstein's view, however, would expect to satisfy.

\medskip
Bell's inequality has been tested in various practical experiments and always found violated. So the currently held belief in physics is that Einstein was wrong\footnote{see, \emph{e.g.}, the account in Wikipedia~\cite{wikipedia}}. Yet, Bell's proposed solution to the EPR objection keeps posing a philosophical puzzle. Models of hidden states and extensions of classical statistics to a theory of \emph{quantum statistics} have been developed in order to solve it (see, \emph{e.g.}, \cite{Bohm-Hiley93}, \cite{Bub76}, \cite{B-NGJ,Gill14}).

\medskip
Indeed, things may be not so clear. The derivation of Bell's inequality assumes that a set of pairwise commuting Heisenberg observables, based on a common deterministic property of a state, should behave like the marginal processes of an underlying (classical) stochastic variable. A violation of the inequality would thus be evidence  that such a hidden state cannot exist.

\medskip
While Bell's theoretical assumption can be justified in the standard Hilbert space model of quantum theory,  conclusions from pratical experiments have to be taken with grains of salt since the experiments are conducted in the environment of Minkowski spaces of relativity theory and not necessarily in pure Hilbert spaces. So Heisenberg measurements need to be studied not only in Hilbert but also in Minkowski space.

\medskip
The present note provides an extended theoretical model for Heisenberg measurements on (finite-dimensional) \emph{relativistic} quantum systems. Such systems do not refer to special physical systems but to the representation of system states as spacelike events in Minkowski space or, more generally, in orthogonal geometries. Hilbert spaces model precisely those orthogonal geometries in which the density function of a Heisenberg measurement is guaranteed to be positive semidefinite (which justifies Bell's assumption for his inequality). All other geometric environments (and thus Minkowski spaces in particular), admit density functions with possibly negative values.

\medskip
So probabilistic conclusions from Heisenberg measurements must be handled with care. Their interpretation in abstract Hilbert space may be subject to larger eigenvalue bounds in Bell's inequality than in Minkowski space. In fact, the Bell bound depends not only on the eigenvalues of the Heisenberg measuring matrix but also on the geometric en\-vironment of the measurement. A measurement with a complete set of observables may violate the (stricter) Bell inequality in Minkowski space while it complies to the (larger) Bell bound in Hilbert space. In fact, we present a numerical example of such a situation in Section~\ref{sec:Bell} below.

\medskip
Each physical measurement constitutes an interaction between the measuring de\-vice and the object to be measured and thus depends on both the properties of the measuring device and on the properties of the object (\emph{cf.}  Section~\ref{sec:interaction}). Seeming randomness of measurement observations could either reflect incomplete information on part of the observer or an inherent randomness on part of the object (\emph{cf.} \cite{Gill-et-al-03}). Or both. The issue appears open to individual interpretations\footnote{\emph{cf.} Schopenhauer~\cite{Schopenhauer}: \flqq Die Welt ist meine Vorstellung \frqq ~(\emph{The world is what I imagine}).}.

\section{Mathematical preliminaries}\label{sec:preliminaries}
This section collects standard facts from linear algebra about complex vector spaces and matrices (see, \emph{e.g.}, \cite{Nering}).
$\R$ denotes the field of real and $\C$ the field of complex scalars. As usual, $\im\in \C$ is the imaginary unit with the property $\im^2 = -1$. $\ov{z}=a-\im b$ is the conjugate of the complex scalar $z=a+\im b$ with $a,b\in \R$. Where $x\in \C^n$ is an $n$-dimensional parameter vector with coordinates $x_i$, $\ov{x}$ denotes the vector with coordinates $\ov{x}_i = \ov{x_i}$. We think of an element $x\in \C^n$ typically as a column vector and write  $x^*=\ov{x}^T$ for the row vector with the coefficients $x_i^* = \ov{x_i}$.

\medskip
$\C^n$ becomes a Hilbert space (\emph{cf.} Ex.~\ref{ex.Hilbert-space} below) under the scalar product
$$
     \(x|y\) = x^*y = \ov{x_1}y_1 + \ldots + \ov{x_n}y_n
$$
or, equivalently, the norm
$$
     \|x\|_2 = \sqrt{|x_1|^2 +\ldots + |x_n|^2} = \sqrt{\(x|x\)}.
$$

\medskip
The \emph{adjoint} $C^*$ of a $n\times n$ matrix $C$ is the transpose of its conjugate:
$$
C^* = \ov{C}^T  \quad\mbox{with the coefficients $C^*_{ij} = \ov{C_{ji}}$.}
$$
In the case $C^*=C$, $C$ is \emph{hermitian} (or \emph{self-adjoint}). The $n\times n$ matrix $U$ is \emph{unitary} if $U^* = U^{-1}$, which yields the identity matrix $I = U^*U$. $$
D=\diag(d_1,\ldots,d_n)
$$ 
denotes a diagonal matrix with diagonal elements $D_{ii} = d_i$. Diagonal matrices with real coefficients are hermitian.

\medskip
\begin{theorem}[Spectral theorem]\label{t.spectral-theorem} The following statements are equivalent:
\begin{enumerate}
\item[(1)] The complex matrix $A$ is hermitian.
\item[(2)] There exist real numbers $\lambda_1,\ldots,\lambda_n$ and a unitary matrix $U$ with column vectors $U_i$ such that
    $$
    A x = \sum_{i=1}^n \lambda_i \(x|U_i\) U_i \quad\mbox{for all $x\in \C^n$.}
    $$
\end{enumerate}
\end{theorem}

\section{Relativistic quantum systems}\label{sec:relativistic-qsystems}
We assume an $n$-dimensional \emph{geometric space} $\cG$, \emph{i.e.}, an $n$-dimensional vector space $\cV$ with the complex field $\C$ of scalars, endowed with a (hermitian) \emph{scalar product} (or \emph{metric}) $(\vx|\vy)$ such that for all $\vx,\vy,\vz\in \cV$ and scalars $\lambda\in \C$,
\begin{eqnarray*}
(\vy|\vx) &=& \ov{(\vx|\vy)}\\
(\vx|\vy +\lambda\vz) &=& (\vx|\vy\) + \lambda(\vx|\vz).
\end{eqnarray*}
Vectors $\vx,\vy\in \cG$ are \emph{orthogonal} if $(\vx|\vy) =0$.  The associated \emph{quadric norm} on $\cG$ is the real-valued function
$$
  \|\vx\|^2 = (\vx|\vx) \quad(\vx\in \cV).
$$
If $\|\vx\|^2 \geq 0$, we refer to $\|\vx\| = \sqrt{\|\vx\|^2}$ as the \emph{length} of $\vx$ in $\cG$. In view of $\|\lambda \vx\| = \lambda\|\vx\|$ for all real $\lambda \geq 0$, the \emph{characteristic surface}
$$
   \cS = \{\vx\in \cG\mid (\vx|\vx) =1\}
$$
represents the $1$-normalized elements of $\cG$ of positive length. A system $\mathfrak S$ is called a \emph{relativistic quantum system} if its states are represented by the elements of the characteristic surface $\cS$ of a geometric space $\cG$.

\subsection{Signed decomposition}\label{sec:signed-decomposition}
Relative to the metric of the geometric space $\cG$, the Spectral Theorem guarantees the existence of a basis $\vE=\{\ve_1,\ldots,\ve_n\}$ and indices $0\leq r\leq s\leq n$ such that
$$
(\ve_i|\ve_j) = \left\{\begin{array}{cll} 0 &\mbox{if $i\neq j$ or $i>s$,}\\
+1 &\mbox{if $i=j$ and $i\leq r$,}\\
-1 &\mbox{if $i=j$ and $r+1\leq i\leq s$.}
\end{array}\right.
$$
and hence
$$
\|\vx\|^2 = |x_1|^2 +\ldots +|x_r|^2 -|x_{r+1}|^2 -\ldots -|x_s|^2,
$$
where $x\in \C^n$ is the coordinate vector of $\vx$ relative to $\vE$. Consequently, $\cG$ decomposes naturally into the sum
$$
 \vG = V^+ \oplus V^- \oplus V^0
$$
of the pairwise orthogonal subspaces
\begin{eqnarray*}
V^0 &=& \span \{\ve_i| s < i\leq n\}\\
V^+ &=& \span \{\ve_i| 0 < i \leq r\}\\
V^- &=& \span \{\ve_i| r < i\leq  s\}.
\end{eqnarray*}

\medskip
Assuming a non-empty characteristic surface $\cS\neq \emptyset$, we have $V^+\neq \emptyset$. In view of
$$
      \cS \subseteq V^+ \oplus V^-,
$$
we furthermore assume $s=n$ without loss of generality. So for every $\vx\in \cG$, there are unique vectors $\vx^+\in V^+$ and $\vx^- \in V^-$ such that
$$
    \vx = \vx^+ +\vx^- \quad\mbox{and} \quad\|\vx\|^2 = \|\vx^+\|^2 + \|\vx^-\|^2.
$$

\medskip
\begin{ex}[Hilbert spaces]\label{ex.Hilbert-space} A \emph{Hilbert space} $\cH$ is a geometric space with $V^+ = \cV$ (\emph{i.e.}, $r=n$). All elements of a Hilbert space have a nonnegative quadric norm
$$
\|\vx\|_2^2 = |x_1|^2 +\ldots + |x_n|^2,
$$
where the $x_i$ are the coordinates of $\vx$ relative to the basis $\vE$. The scalar product of a Hilbert space is denoted by
$$
\(\vx|\vy\) = \ov{x_1}y_1 +\ldots + \ov{x_n}y_n.
$$

\medskip
For a general geometry $\cG$ with characteristic subspaces $V^+$ and $V^-$, the restriction $\cG^+$ of $\cG$ to $V^+$ is a Hilbert space, while $V^-$ becomes a Hilbert space relative to the negative of the scalar product:
$$
    \(\vx|\vy\) = -(\vx|\vy).
$$
Hence the quadric norm of $\cG$ can be expressed in terms of the Hilbert norm:
\begin{equation}\label{eq.signed-quadric-norm}
\|\vx\|^2 = \|\vx^+\|^2 + \|\vx^-\|^2  = \|\vx^+\|^2_2 - \|\vx^-\|^2_2.
\end{equation}
\end{ex}

\medskip
\begin{ex}[Minkowski spaces]\label{ex.Minkowski-space} An ($n$-dimensional)  \emph{Minkowski space} $\cM$ is a geometric space with $r=n-1$ and thus the quadric norm
$$
  \mu(\vx) = |x_1|^2 + \ldots + |x_{n-1}|^2 - |x_n|^2.
$$
Minkowski spaces yield models for relativity theory where elements $\vx\in \cM$ are time tagged \emph{events}. Events $\vx$ with positive norm $\mu(\vx)> 0$ are \emph{spacelike}. For any $\vx\in \cM$, the affine subspace
$$
   \cM(\vx) = \{\vy\in \cM\mid y_n = x_n\}.
$$
consists of the events with the same time tag $x_n$.
\end{ex}

\subsection{Isometries, Galilei and Lorentz transformations}\label{sec:isometries}
An \emph{isometry} of the geometry $\cG$ is a metric preserving linear operator $T:\cG\to \cG$:
$$
    (T\vx|T\vy) = (\vx|\vy\) \quad\mbox{holds for all $\vx, \vy\in \cG$}.
$$

\medskip
\begin{Bemerkung}
An isometry $T$ corresponds to a transition to a new reference basis
$$
\vE' = T(\vE) =\{T\ve_1,\ldots, T\ve_n\} =\{\ve_1',\ldots,\ve_n'\}
$$
that exhibits the same geometric structure.
\end{Bemerkung}

\medskip
The isometries of the Hilbert space $\cH$ are the so-called \emph{unitary} operators. If the scalars of $\cH$ are  restricted to the real coefficient field $\R$, its isometries are also known as \emph{Galiliei transformations}.  Unitary operators yield typical examples of isometries in a general geometry $\cG$ as follows:

\medskip
Let $U^+$ and $U^-$ be unitary operators on $V^+$ and $V^-$ respectively, then
$$
 U\vx = U^+\vx^+ + U^-\vx^-
$$
is a unitary isometry of $\cG$.

\medskip
Isometries of Minkowski spaces are \emph{Lorentz transformations}.  A Lorentz trans\-for\-mation of $\cM$ that fixes the last (time) coordinate, for example, arises from a unitary operator on the $n-1$ dimensional Hilbert space $V^+$ in this fashion.

\section{Heisenberg measurements}\label{sec:Heisenberg-measurements}
Consider a relativistic quantum system $\cS$ in a geometry $\cG$ and a $n\times k$ matrix $W$ with rows $W_i$ and coefficients $W_{ij} \in \R$. $W$ is the \emph{eigenvalue matrix} of the measuring instrument to be defined. Any $\vx\in \cG$ with coordinates $x_i$ relative to the basis $\vE$ decomposes into a sum of pairwise orthogonal vectors $\vx_w$:
\begin{equation}\label{eq.eigenvalue-decomposition}
\vx = \sum_{w}\vx_w  \quad\mbox{with} \quad \vx_w = \D\sum_{W_i =w} x_i\ve_i.
\end{equation}
With the eigenvalue decomposition (\ref{eq.eigenvalue-decomposition}) of $\vx$ one associates a
\emph{density} as the function
$$
q_\vx(w) = (\vx_w|\vx_w) =\|\vx_w\|^2
$$
with the property
$$
    \int_{\R^k} dq_\vx(w) = \sum_w q_\vx(w) = 1 \quad\mbox{for all $\vx\in \cS$},
$$
which yields
$$
W(\vx) = \int_{\R^k} w~d q_\vx(w) = \sum_{i=1}^r W_i|x_i|^2 - \sum_{j=r+1}^n W_j|x_j|^2.
$$

\medskip
\begin{Bemerkung} A density is a \emph{signed measure} in the sense of mathematical measure theory\footnote{see, \emph{e.g.}, \cite{Cohn97}}. It can always be expressed as the difference of two nonnegative measures, in complete analogy with (\ref{eq.signed-quadric-norm}).
\end{Bemerkung}

\medskip
A \emph{Heisenberg measuring instrument} on $\cS$ is a pair $H(W,T)$ with an eigenvalue matrix  $W$ as above and $T$ an isometry of $\cG$. $H(W,T)$  produces $k$ simultaneously measured real para\-meters as the components of the vector
$$
    h(\vx) = W(T\vx) = (h_1(\vx),\ldots,h_k(\vx)) \in \R^k
$$
when $\cS$ is in the state $\vx$.

\medskip
The $h_j(\vx)$ are single-valued Heisenberg measurements of the form $H(W_j,T)$ in their own right, each with respect to the same isometry $T$. In this sense, the marginal Heisenberg measurement functions
$$
   \vx \mapsto h_j(\vx) \quad (j=1,\ldots,k)
$$
constitute a \emph{complete set of common observables} on the relativistic quantum system $\cS$.

\medskip
\begin{Bemerkung} Note that the eigenvalue decomposition of $\vx$ relative to $W$ is a common refinement of the eigenvalue decompositions of $\vx$ relative to the $k$ component measurements.
\end{Bemerkung}

\subsection{Matrix representations}
With reference to the basis $\vE =\{\ve_1,\ldots,\ve_n\}$ of $\cV$, one may identify any $\vx\in \cV$ with its vector $x\in \C^n$ of $\vE$-coordinates $x_i$. In particular, an isometry $T$ of $\cG$ becomes a $n\times n$ matrix. Letting
$$
    \Lambda^{(j)} = \diag(W_{1j},\ldots,W_{nj}) \quad(j=1,\ldots,k),
$$
the $k$ marginal measurements of the Heisenberg measuring instrument $H(W,T)$ can then be expressed in matrix notation as
\begin{equation}\label{eq.self-adjoint-representation}
h_j(x) = x^* A^{(j)} x \quad\mbox{with $A^{(j)}= T^*\Lambda^{(j)}T$.}
\end{equation}

\medskip
Note that the matrices $A^{(j)}$ are hermitian, which means that single-valued Heisenberg measurements in $\cG$ are also Heisenberg measurements in the the standard model of quantum theory.

\medskip
Moreover, if $T$ is a unitary matrix (\emph{i.e.}, if $T^{-1} = T^*$), then the measurement matrices $A^{(j)}$ commute pairwise:
\begin{equation}\label{eq.Heisenberg-commutativity}
A^{(j)} A^{(\ell)} = A^{(\ell)}A^{(j)} \quad\mbox{for all $j,\ell =1,\ldots, k$}.
\end{equation}

\medskip
\begin{Bemerkung} By the Spectral Theorem,  a matrix $A$ is hermitian if and only if there exist a real dia\-gonal matrix $\Lambda$ and a matrix $T$ such that $A=T^*\Lambda T$. $T$ does not have to be an isometry.
\end{Bemerkung}

\medskip
\begin{ex}[Observables in Hilbert space] A set $\cA=\{A^{(1)},\ldots, A^{(k)}\}$ of $k$ pairwise commuting hermitian matrices $A^{(j)}$ admits a representation of type (\ref{eq.self-adjoint-representation}) with a common unitary matrix $T$ and suitable real diagonal matrices $\Lambda^{(j)}$.

\medskip
Any isometry $T$ of a Hilbert space is described by a unitary matrix. Consequently, the commutativity condition (\ref{eq.Heisenberg-commutativity}) characterizes complete sets of common observables in Hilbert space.
\end{ex}

\subsection{Stochastic interpretations}
Say that the Heisenberg measurement $H(W,T)$ is \emph{positive semidefinite} in the relativistic quantum state  $\vs\in \cS$ if the associated density $q_{T\vs}$ is nonnegative and, therefore, a probability distribution. In this case, the measurement can be interpreted as the expected value
$$
   E(X) = \int_{\R^k} w~d q_{\vx}(w) = W(\vx) \quad\mbox{with $\vx = T\vs$}
$$
of a stochastic variable $X$ with the probability distribution
$$
\Pr\{X = w\} = q_{\vx}(w).
$$

\medskip
\begin{ex} Heisenberg measurements in Hilbert space are positive semidefinite in any quantum state and, therefore, always allow the interpretation of the measurement result as an expected value of  eigenvalues.
\end{ex}

\medskip
In fact, every Heisenberg measurement admits a stochastic interpretation. At least in principle. To this end, make the eigenvalue matrix $W$ depend on the state $\vx$ and define $W^\vx$ as the modified matrix with rows
$$
   W^{\vx}_{i}  = \left\{\begin{array}{rll} \|\vx\|_2^2 W_{i} &\mbox{if $i\leq r$}\\
   -\|\vx\|^2_2 W_{i} &\mbox{if $i > r$.}\end{array}    \right.
$$
If $\vx\neq \v0$,  the parameters $p_i(\vx) = |x_i|^2/\|\vx\|_2^2$ form a probability distribution and yield
\begin{equation}\label{eq.stochastic-interpretation}
    W(\vx) = \sum_{i=1}^n W^\vx_i p_i(\vx)
\end{equation}
as the corresponding expected value of the rows of $W^\vx$.

\medskip
\begin{Bemerkung} While the stochastic model (\ref{eq.stochastic-interpretation}) shows that Heisenberg measurements on relativistic quantum systems may be analyzed with standard methods of mathematical statistics, it is of problematic use if one wants to check experimentally theoretically predicted properties which depend on the {\it a priori} eigenvalues of the measuring instrument.
\end{Bemerkung}

\subsection{Measurement functions and hidden states}\label{sec:hidden-states}
An $n$-dimensional \emph{hidden state model} of a system $\mathfrak S$ assumes the existence of a set $$
\Omega=\{\omega_1,\ldots,\omega_n\}
$$
of $n$ definite (but possibly ''hidden'') ground states $\omega_i$. In this context,  a \emph{measurement function} is just a function
$$
X:\Omega \to \R.
$$
A \emph{(general) state} of $\mathfrak S$ is a linear superposition of ground states and thus corresponds to a coefficient vector $s\in \C^n$. Within appropriate geometries, these coefficient vectors obey certain metrics. The particular superposition of the ground states is thought to govern stochastic aspects of the measurement {\it via} $X$. Thus, general states are typically normalized to length $\|s\|=1$ relative to the relevant geometry.

\medskip
\begin{ex}\label{ex.quantum-computing} The standard model of \emph{quantum computation} (\emph{cf.} \cite{Nielsen-Chuang}) is a hidden state model. It assumes a Hilbert space as its geometric environment.
\end{ex}

Feynman~\cite{Feynman87} (see also \cite{Scully-et-al-94}) describes a spin $1/2$ system with $4$ ground states $\omega_1,\omega_2,\omega_3,\omega_4$. Assume that $X$ and $Y$ measure the spin around the $x-$ and the $y-$axis relative to the ground states with the returns
\begin{equation}\label{eq.Feynman}
  \begin{array}{c|cccccc}
    &\omega_1&\omega_2&\omega_3&\omega_4 \\  \hline
    X &-1&-1&+1&+1\\
    Y &-1&+1&-1&+1,
  \end{array}
\end{equation}
which corresponds to the Heisenberg measurement matrices
\begin{eqnarray*}
A &=& \diag(-1,-1,+1,+1)\\
B&=&\diag(-1,+1,-1,+1).
\end{eqnarray*}
It can be argued that the following observation probabilities are possible:
$$
\begin{array}{ccccccccc}
\Pr\{X= +1\} &=& 1/4 &, &\Pr\{X=-1\}&= &3/4\\
\Pr\{Y= +1\}&=& 0 &,&\Pr\{Y=-1\}&=& 1.
\end{array}
$$

It is easy to see that there is no $4$-dimensional Hilbert space quantum state that would imply these probabilities. However, the relativistic quantum state
$$
     s =(\sqrt{5/8}, \sqrt{1/8},\sqrt{3/8}, \sqrt{1/8})
$$
in $4$-dimensional Minkowski space (with metric $(x|y)$) is positive semidefinite for the eigenvalue matrix
$$
    W = \begin{bmatrix} -1 &-1&-1&+1 \\
    -1&+1&-1&+1\end{bmatrix}^T .
$$
So $X$ and $Y$ are jointly observable in the (Minkowski) state $s$ with the probabilities as above and the expected values
$$
E(X) = (s|As) = -1/2 \quad\mbox{and}\quad E(Y) = (s|B) = -1.
$$

Since $\|s\|_2 = \sqrt{5}/2$, an interpretation as a joint measurement in Hilbert space according to (\ref{eq.stochastic-interpretation}) would re-scale $s$ to the vector 
$$
     \tilde{s} = \frac{s}{\|s\|_2} = \frac{2s}{\sqrt{5}} \quad\mbox{of Hilbert norm}\quad \|\tilde{s}\|_2 = 1. 
$$
With the re-scalded measurement matrices
\begin{eqnarray*}
A'' &=& \diag(-\sqrt{5}/2,-\sqrt{5}/2,+\sqrt{5}/2,-\sqrt{5}/2)\\
B''&=&\diag(-\sqrt{5}/2,+\sqrt{5}/2,-\sqrt{5}/2,-\sqrt{5}/2).
\end{eqnarray*}
one then computes relative to $\tilde{s}$ in Hilbert space: 
$$
\(\tilde{s}|A''\tilde{s}\) =(s|As) = E(X) \quad\mbox{and}\quad \(\tilde{s}|B''\tilde{s}\) =(s|Bs) = E(Y).
$$

\subsection{Interactions and Schr\"odinger type evolutions}\label{sec:interaction}
An $n$-dimensional \emph{interaction system} assumes patterns of pairwise interactions of $n$ particles $i,j$ that are described by (real) \emph{interaction coefficients} $J_{ij}$. An interaction pattern is thus represented by an $n\times n$  real matrix $J$. $J$ has a unique decomposition
$$
   J = J^{(0)} + J^{(1)}
$$
into a symmetric matrix $J^{(0)}$ and a skew-symmetric matrix $J^{(1)}$, which gives rise to a (complex) hermitian matrix
\begin{equation}\label{eq.interaction-matrix}
  \hat{J} = J^{(0)} + \im J^{(1)}.
\end{equation}
In fact, (\ref{eq.interaction-matrix}) establishes an isomporphism between the vector space of all real $n\times n$ matrices and the space of all hermitian $n\times n$ matrices (over the real field $\R$). So any hermitian  $n\times n$ matrix $A$ reflects the superposition of a symmetric and a skew-symmetric interaction pattern of $n$ particles. Consequently,
$$
   x^*Ax = \sum_{i,j} A_{ij}\ov{x_i} x_j
$$
is the total value of interaction if the particles $i$  interact with intensity levels $x_i$ according to the pattern represented by $A$.

\medskip
Let $T$ be an isometry of the geometric space $\cG$ and consider for the relativistic state $\vx\in \cS$  the sequence of values
\begin{equation}\label{eq.Schroedinger-evolution}
   \varphi_t(\vx) = x^*[(T^t)^*AT^t]x = (T^tx)^* A(T^tx) \quad (t=0,1,\ldots)
\end{equation}

This sequence admits interpretations that are dual to each other. The \emph{Schr\"odinger picture} suggests a constant interaction pattern $A$ relative to an evolution of states
$$
\vx_0, \vx_1,\ldots, \vx_t, \vx_{t+1} \ldots \quad\mbox{with $\vx_{t+1} = T\vx_t$.}
$$
The \emph{Heisenberg picture}, on the other hand, sees an evolution of interaction patterns relative to a constant state $\vx$:
$$
    A_0, A_1, \ldots, A_t, A_{t+1},  \ldots  \quad\mbox{with $A_{t+1} = T^*A_tT$.}
$$

\medskip
Either way, and without any assumptions on a probabilistic nature of a Heisenberg measurement, one can derive (\emph{cf.} \cite{Faigle-Schoenhuth}) the analogue of \emph{von Neumann's theorem}:
\begin{itemize}
\item[($\bullet$)] \emph{The sampling averages of the $\varphi_t(\vx)$ converge if the Hilbert norm of $T$ is bounded.}
\end{itemize}

\medskip
\begin{Bemerkung} The idea of pairwise interaction is fundamental in a wide variety of mathematical application models (\emph{cf.} \cite{Faigle22}).
\end{Bemerkung}

\section{The Bell inequality}\label{sec:Bell}
We first note an inequality that goes back to Bell~\cite{Bell64} for sets of sets of three common Heisenberg observables. We define the \emph{Bell number} of  the corresponding eigenvalue matrix $W\in \R^{n\times 3}$ as its maximum norm
$$
   |W|_\infty = \max_{ij} |W_{ij}|.
$$
Let  $\vs$ be a member of some  $n$-dimensional geometry $\cG$ with quadratic norm
$$
\|\vs\|^2 = \sum_{i=1}^n g_i |s_i|^2 =1 \quad\mbox{where}\quad g_i\in \{-1,+1\}
$$
and the associated density relative to $W$: 
$$
   q_\vs(x,y,z) = \sum_{W_i=(x,y,z)} g_i|s_i|^2.
$$
We consider the measurement values
\begin{eqnarray*}
W(XY) &=& \sum_{W_i =(x,y,z)} x y~g_i|s_i|^2  =  \sum_{(x,y,z)\in \R^3} xy~q_\vs(x,y,z)\\
W(YZ) &=& \sum_{W_i =(x,y,z)} yz~g_i|s_i|^2  =  \sum_{(x,y,z)\in \R^3} yz~q_\vs(x,y,z)\\
W(XZ) &=& \sum_{W_i =(x,y,z)} xz~g_i|s_i|^2  =  \sum_{(x,y,z)\in \R^3} xz~q_\vs(x,y,z) .
\end{eqnarray*}

\begin{lemma}\label{l.hilfs-Bell} If the density $q_\vs$ is nonnegative, then the measurements satisfy the inequality
\begin{equation}\label{eq.Bell-expectation}
|W(XY) -W(YZ)| + W(XZ) \;\leq\; |W|_\infty^2.
\end{equation}
\end{lemma}

\Pf Any row triplet $(x,y,z)\in W$ satisfies the inequality
$$
|xy -yz| +xz \;\leq\; W^2_\infty.
$$
If $q_\vs$ is nonnegative, and hence a probability distribution on the $n$ rows of $W$, the measurement values are the corresponding expectations of the component products, which implies
$$
  |W(XY) -W(YZ)| + W(XZ) \leq  W(|XY - YZ| +XZ) \;\leq\; W^2_\infty.
$$
\qed

\medskip
Lemma~\ref{l.hilfs-Bell} immediately yields:
\begin{theorem}[Bell inequality]\label{t.Bell} Let $W\in \R^{n\times 3}$ be an eigenvalue matrix of a Heisenberg measurement and $\vx\in \cS$ a relativistic quantum state with a nonnegative density function $q_\vx$. Let $X,Y,Z$ be the associated stochastic component variables so that
$$
W(\vx) = \int_{\R^3} w~d q_\vx(w) = (E(X),E(Y),E(Z)).
$$
Then the expected values of the pairwise products satisfy the inequality
\begin{equation}\label{eq.big-Bell}
|E(XY) -E(YZ)| + E(XZ) \;\leq\;  |W|_\infty^2 .
\end{equation}
\end{theorem}

\medskip
For an illustration of the Bell inequality in the Minkowski space $\M_5$ relative to the coordinate space $\C^5$,  consider the three hermitian and pairwise commuting measurement matrices
\begin{eqnarray*}
A &=&\diag(-1,+1,+1,+1,+1)\\
B & =&\diag(-1,+1,-1,-1,+1)\\
C & =&\diag(-1,-1,-1,+1,+1)
\end{eqnarray*}
with the Bell bound $|W|_\infty=1$ on the eigenvalues. For any $x\in \cM$, one has
$$
    (Ax)_5 = (Bx)_5 = (Cx)_5 =x_5.
$$
So an associated Heisenberg measurement $H$ with the identity $T=I$ as the trivial iso\-metry and com\-ponents
$$
 h(x) = (x^*Ax, x^*Bx, x^*Cx)
$$
takes place in the $4$-dimensional subspace $\M_5(x)$ of events with the same time tag. There is no \emph{time loophole} to be considered.  For a concrete example, fix
$$
    s = \frac{\sqrt{3}}{3}(1,1,1,1,1).
$$
Because $\mu(s) = 1$, $s$ represents a relativistic quantum state in $\M_5$. Moreover, each {\em pair} of measurements is positive semidefinite in the state $s$. For the expected values of the pairwise products of the cor\-responding stochastic variables $X,Y,Z$, one computes
$$
E(XY) =-1/3, \; E(XZ) =+1, \;E(YZ) =-1,
$$
which violates the Bell inequality however:
$$
|E(XY) -E(YZ)| + E(XZ) = 2/3 + 1 \;>\; 1 =|W|_\infty^2 .
$$
This shows that the hypothesis of positive semidefinitenes of the joint measurement of \emph{all} three measurements in Theorem~\ref{t.Bell} cannot be dropped in general. $h$ represents a complete set of three common observables. Yet, $h$ is not positive semidefinite for the \emph{complete} set in the state $s$.

\medskip
In view of $\|s\|_2^2 = 5/3$, on the other hand, the stochastic Hilbert space measurement model (\ref{eq.stochastic-interpretation}) refers to the measurement matrices
\begin{eqnarray*}
A'' &=&\diag(-5/3,+5/3,+5/3,+5/3,-5/3)\\
B'' & =&\diag(-5/3,+5/3,-5/3,-5/3,-5/3)\\
C'' & =&\diag(-5/3,-5/3,+5/3,+5/3,-5/3)
\end{eqnarray*}
with the Bell number $|W''|_\infty=5/3$. For the corresponding stochastic variables $X''$, $Y''$, $Z''$, one obtains
$$
E(X'' Y'') = E(X''Z'') = E(Y''Z'') = 25/9,
$$
which satisfies the pertaining Bell inequality
$$
|E(X''Y'') -E(Y''Z'')| + E(X''Z'') = 25/9 =|W''|^2_\infty .
$$

\medskip
\begin{Bemerkung} The example in the present section shows that a seeming violation of the Bell inequality may result from the stochastic interpretation of the measurement model and the geometric environment in which the measurement is carried out. It is not necessarily an indication of an underlying physical reality {\it per se}.
\end{Bemerkung}

\section{Final Remarks}
The facts that the Minkowski norm may take on negative values and that densities are signed measures makes it natural to study physical phenomena with a stochastic appearence in mathematical models where the descriptive parameters can be ''negative probabilities''. Although still not standard in quantum theory, this modeling approach was already taken by Wigner~\cite{Wigner32} and recommended by Feynman~\cite{Feynman87} (see also \cite{BlassGurevich21,Scully-et-al-94}, for instance).

\medskip
A fundamental model in applied statistics assumes that stochastic phenomena are sequentially observed according to a Markov chain with internal and possibly hidden states (\emph{cf.} (\cite{EphraimMerhav02}). Markov chains do not appear to fit under the umbrella of Schr\"odinger quantum state evolutions. Permitting ''negative probabilities'', however, one is lead to a unifying theory of statistical evolutions.

\medskip
The statistical model \cite{FaigleGierz}, for example, admits stochastic processes that are described by evolutions under the action of Riesz operators on Banach spaces. Riesz measurements generalize Heisenberg measurements. Moreover, mean ergodic evolutions can be characterized in this generality. Classical Markov chains and quantum random walks (\emph{cf.} \cite{Temme-et-al-11}) as well as (discrete) Schr\"odinger evolutions fit into this context as special cases. Similarly, the theory of quantum information and computing (\emph{cf.} \cite{Nielsen-Chuang}) appears to extend accordingly. (The algebraic model \cite{FroehlichPizzo22} for a statistical generalization of Schr\"odinger evolutions seems to aim into a different direction.)

\end{document}